\documentstyle[aps,prl,multicol,epsf]{revtex}
\begin{document}
\newcommand{\tbox}[1]{\mbox{\tiny #1}}
\newcommand{\half}{\mbox{\small $\frac{1}{2}$}}
\newcommand{\mbf}[1]{{\mathbf #1}}

\title{Wavepacket dynamics in energy space, 
RMT and quantum-classical correspondence}

\author
{
Doron Cohen$^1$, Felix M. Izrailev$^2$ and Tsampikos Kottos$^3$\\
\footnotesize
$^1$
Department of Physics, Harvard University, Cambridge, MA 02138 \\
$^2$
Instituto de Fisica, Universidad Autonoma de Puebla,
Puebla, Pue 72570, Mexico\\
$^3$
Max-Planck-Institut f\"ur Str\"omungsforschung,
37073 G\"ottingen, Germany
}

\date{November 1999, to be published in Phys. Rev. Lett.}

\maketitle


\begin{abstract}
We apply random-matrix-theory (RMT) to the analysis of evolution of wavepackets
in energy space. We study the crossover from ballistic behavior to saturation, 
the possibility of having an intermediate diffusive behavior, 
and the feasibility of strong localization effect. Both theoretical 
considerations and numerical results are presented. 
Using quantal-classical correspondence (QCC) considerations we question the 
validity of the emerging dynamical picture.  In particular we claim 
that the appearance of the intermediate diffusive behavior is possibly 
an artifact of the RMT strategy.  
\end{abstract}

\begin{multicols}{2}

We are interested in the dynamics that is generated by a Hamiltonian
of the type ${\cal H}=\mbf{E}+\mbf{W}$ where $\mbf{E}$ is a diagonal
matrix whose elements are the ordered energies $\{E_n\}$, with mean
level spacing $\Delta$, and $\mbf{W}$ is a banded matrix. It is assumed
that $\mbf{W}$ is similar to a `Banded Random Matrix' (BRM), with
non-vanishing couplings within the band $0 < |n-m| \le b$. These
coupling elements are zero on the average, and they are characterized
by the variance $\sigma=(\langle |\mbf{W}_{nm}|^2 \rangle)^{\tbox{1/2}}$.
Thus, there are four parameters $(\Delta,b,\sigma,\hbar)$ that controls 
the dynamics.     
One important application of BRM is in solid-state physics for the
study of localization in quasi-one-dimensional disordered systems.
In this frame non-zero values of $\Delta$ reflect the presence
of a constant electric field along the sample.
However, in this Letter we mainly have in mind the original motivation
following Wigner \cite{wigner}. Namely, the study of either {\em chaotic} 
or complex conservative quantum systems that are encountered
in nuclear physics as well as in atomic and molecular physics. For this
reason the above defined model (with non-zero $\Delta$) is known in the
literature \cite{mario,flamb,felix1} as Wigner's BRM (WBRM) model.

Consider a system whose total Hamiltonian is ${\cal H}(Q,P)$, 
where $(Q,P)$ is a set of canonical coordinates. 
We assume that the preparation and the representation 
of the system are determined by a Hamiltonian ${\cal H}_0(Q,P)$. 
We also assume that both ${\cal H}_0(Q,P)$ and ${\cal H}(Q,P)$ generate 
classically chaotic dynamics of similar nature$^{\dag}$. 
We choose a basis such that 
the quantized Hamiltonian matrix ${\cal H}_0$ has a diagonal
structure ${\cal H}_0 = \mbf{E}$. According to general semiclassical 
arguments \cite{mario}, the quantized Hamiltonian matrix ${\cal H}$,  
in the same basis, has a band structure ${\cal H} = \mbf{E} + \mbf{W}$. 
The WBRM model can be regarded as a {\em simplified} local description 
of the true Hamiltonian matrix. However, there is one feature that 
distinguishes the effective WBRM model from the true Hamiltonian. It is 
the assumption that the off-diagonal elements are {\em uncorrelated}, 
as if they were independent random numbers. In this Letter 
we would like to explore the consequences of this RMT assumption 
on the dynamics. 
Below we define the classical limit of the WBRM-model, 
and the various parametric regimes in the quantum-mechanical theory. 
We analyze the dynamical scenario in each regime, and we explain 
that the emerging picture is incompatible with the 
quantal-classical correspondence (QCC) principle.

Taking ${\cal H}(Q,P)$ to be a generator for the (classical) dynamics, 
the energy ${\cal H}_0(t)={\cal H}_0(Q(t),P(t))$ fluctuates. 
The fluctuations are characterized by a correlation time
$\tau_{\tbox{cl}}$, and by an amplitude $\delta E_{\tbox{cl}}$.
The three parameters $(\Delta,b,\sigma)$ that define the 
effective WBRM model are determined by semiclassical 
considerations \cite{mario}. One obtains $\Delta\propto\hbar^{d}$, 
and $b\propto\hbar^{-(d-1)}$, and $\sigma\propto\hbar^{(d-1)/2}$,
where $d$ is the number of degrees of freedom (dimensionality) of the system.
In this Letter, we find it convenient to {\em define}$^\ddag$ 
the two classical quantities ($\tau_{\tbox{cl}},\delta E_{\tbox{cl}}$)  
in terms of the common quantum-mechanical parameters:   
\begin{eqnarray} \label{eq1}
\tau_{\tbox{cl}} \ = \ \hbar/(b\Delta)
\ \ , \ \ \ \ \ \ \
\delta E_{\tbox{cl}} \ = \ 2\sqrt{b} \ \sigma
\end{eqnarray}
The classical dynamical scenario is formulated 
by using a phase-space picture \cite{crs}. 
The initial preparation is assumed to be a microcanonical distribution 
that is supported by one of the energy-surfaces of ${\cal H}_0(Q,P)$. 
For $t>0$, the phase-space distribution 
spreads away from the initial surface. `Points' of the evolving
distribution move upon the energy-surfaces of ${\cal H}(Q,P)$.
We are interested in the distribution of the energies ${\cal H}_0(t)$
of the evolving `points'. It is easily argued that for short times 
this distribution evolves in a ballistic fashion. 
Then, for $t \gg \tau_{\tbox{cl}}$, 
due to ergodicity, a `steady-state distribution' appears, 
where the evolving `points' occupies an `energy shell' in phase-space. 
The thickness of this energy shell \cite{felix1} 
equals $\delta E_{\tbox{cl}}$.
Thus we have a crossover from ballistic energy spreading 
to saturation. The dynamics in the classical limit 
is fully characterized by the two classical 
parameters $\tau_{\tbox{cl}}$ and $\delta E_{\tbox{cl}}$.

We are going to study the corresponding quantum-mechanical scenario.
We want to explore the consequences of assuming that WBRM model 
can be used as an effective model for the true dynamics. At $t=0$ the
system is prepared in an eigenstate of ${\cal H}_0 = \mbf{E}$. 
For $t>0$ the evolution of the system is determined by 
${\cal H} = \mbf{E}+\mbf{W}$. The evolving state is $\psi(t)$, 
and we are interested in the evolving 
distribution $|\langle n| \psi(t) \rangle|^2$.
We shall use the following terminology: 
The standard perturbative regime is $(\sigma/\Delta)\ll 1$; 
The Wigner regime is $1 \ll (\sigma/\Delta) \ll b^{\tbox{1/2}}$;
The ergodic regime is $b^{\tbox{1/2}} \ll (\sigma/\Delta) \ll b^{\tbox{3/2}}$;
The localization regime is $b^{\tbox{3/2}} \ll (\sigma/\Delta)$. 
See \cite{wigner,felix1}. It is easily
verified that the limit $\hbar\rightarrow 0$ corresponds to the ergodic
regime or possibly (provided $d=2$) to the localization regime.

The structure of the eigenstates $\alpha$ of ${\cal H}$ has been studied in
\cite{wigner,flamb,felix1}. We denote the average shape of an eigenstate as
$W_{\tbox{E}}(r)=\langle|\varphi_{\alpha}(n_{\alpha}+r)|^2\rangle$ where
$\varphi_{\alpha}(n)=\langle n|\alpha \rangle$, and $n_{\alpha}$ is the
'site' around which the eigenstate is located. The average is taken over all
the eigenstates that have roughly the same energy $E_{\alpha} \sim E$.
There are two important energy scales: One is the classical width of the
energy shell $\delta E_{\tbox{cl}}$, and the other is the range of the
interaction $\Delta_b = b\Delta$.
In the {\em standard perturbative regime} $W_{\tbox{E}}(r)$ contains
mainly one level, and there are perturbative tails that extend over the
range $\Delta_b$. In the {\em Wigner regime}, many levels are mixed:
the main (non-perturbative) component of $W_{\tbox{E}}(r)$ has width
$\Gamma=2\pi (\sigma/\Delta)^2\times\Delta$, and the shape within the
bandwidth $\Delta_b$ is of Lorentzian type. However in actual physical
applications this Lorentzian is a special case of core-tail structure
\cite{crs}, where the tail can be found via first order perturbation
theory. Outside the bandwidth the tails decay faster than exponentially
\cite{flamb}.
On approach to the ergodic regime $W_{\tbox{E}}(r)$ spills over the range
$\Delta_b$. Deep in the {\em ergodic regime} it occupies `ergodically'
the whole energy shell whose width is $\delta E_{\tbox{cl}}$. In actual
physical applications the exact shape is determined by simple classical
considerations \cite{crs,felix2}.
Deep in the {\em localization regime} $W_{\tbox{E}}(r)$ is no longer
ergodic: A typical eigenstate is exponentially localized within an energy
range $\delta E_{\xi} = \xi\Delta$ much smaller than $\delta E_{\tbox{cl}}$.
The localization length is $\xi \approx b^2$. In actual physical applications
it is not clear whether there is such type of localization. To avoid
confusion, we are going to use the term `localization' only in the
sense of having $\delta E_{\xi}\ll\delta E_{\tbox{cl}}$.

Now we would like to explore the various dynamical scenarios that
can be generated by the Schr\"odinger equation for
$a_n(t)=\langle n|\psi(t) \rangle$. Namely,
\begin{eqnarray} \label{eq3}
\frac{da_n}{dt} \ = \
-\frac{i}{\hbar} E_n \ a_n
\ -\frac{i}{\hbar}\sum_{m} \mbf{W}_{nm} \ a_{m}
\end{eqnarray}
starting with an initial preparation $a_n=\delta_{nm}$ at $t{=}0$. 
In a previous study \cite{kottos} only the localization regime 
has been considered. Here we are going to consider the general case
($\Delta\ne 0$). We describe the energy spreading profile for $t>0$
by the transition probability kernel $P_t(n|m)=\langle |a_n(t)|^2 \rangle$.
The angular brackets stand for averaging over realizations of the Hamiltonian.
In particular, it is convenient to characterize the energy spreading profile
by the variance $M(t)=\sum_n (n{-}m)^2 P_t(n|m)$, and by the participation
ratio $N(t)= (\sum_n (P_t(n|m))^2)^{-1}$, and by the total transition
probability $p(t)$, and by the out-of-band transition probability $q(t)$.
Both $p(t)$ and $q(t)$ are defined as $\sum_n' P(n|m)$ where the prime
indicate exclusions of the term $n=m$ or exclusion of the terms $|n{-}m| \le b$
respectively. Equation (\ref{eq3}) has been integrated numerically using the
self-expanding algorithm of \cite{kottos} to eliminate finite-size
effects. Namely, additional $10b$ sites are added to each edge
whenever the probability of finding the `particle' at the edge sites 
exceeded $10^{-15}$. 
Fig.1 illustrates the time-evolution of the energy spreading 
profile. From such plots we can define various time
scales. The times $t_{\tbox{ball}}$ and $t_{\tbox{sat}}$ pertains to $M(t)$
and mark the departure-time from ballistic behavior and the crossover-time
to saturation. The time $t_{\tbox{sta}}$ pertains to $N(t)$ and marks the
crossover to a stationary distribution. The time scale $t_{\tbox{prt}}$
pertains to $p(t)$ and marks the disappearance of the simple perturbative
structure (See (\ref{eq4}) below). The asymptotic value of $q(t)$, if it is
much less than $1$, indicates that the system is either in 
the standard perturbative regime
or in the Wigner regime, where out-of-band transitions can be neglected.
The saturation profile is given by the expression
$P_{\tbox{$\infty$}}(n|m)=\sum_{\alpha} |\langle n|\alpha\rangle|^2
|\langle \alpha|m\rangle|^2$, and it is roughly approximated by the
auto-convolution of $W_{\tbox{E}}(r)$. Therefore the 
saturation profiles (Fig.1) are similar to the average shape of 
the eigenstates. We have found that $M(\mbox{\small $\infty$})$ satisfies a scaling 
relation $Y = 2X{\cdot}(1{-}\exp(-1/(2X))$ where 
$X{=}(\sigma/\Delta)/b^{\tbox{3/2}}$ and $Y{=}(M(\mbox{\small $\infty$}))^{\tbox{1/2}}/b^2$. 
This scaling relation is similar to the one that pertains to
the average shape of the eigenstates \cite{felix1}.

\begin{figure}
\hspace*{-0.5cm}
\epsfysize=1.3in
\epsffile{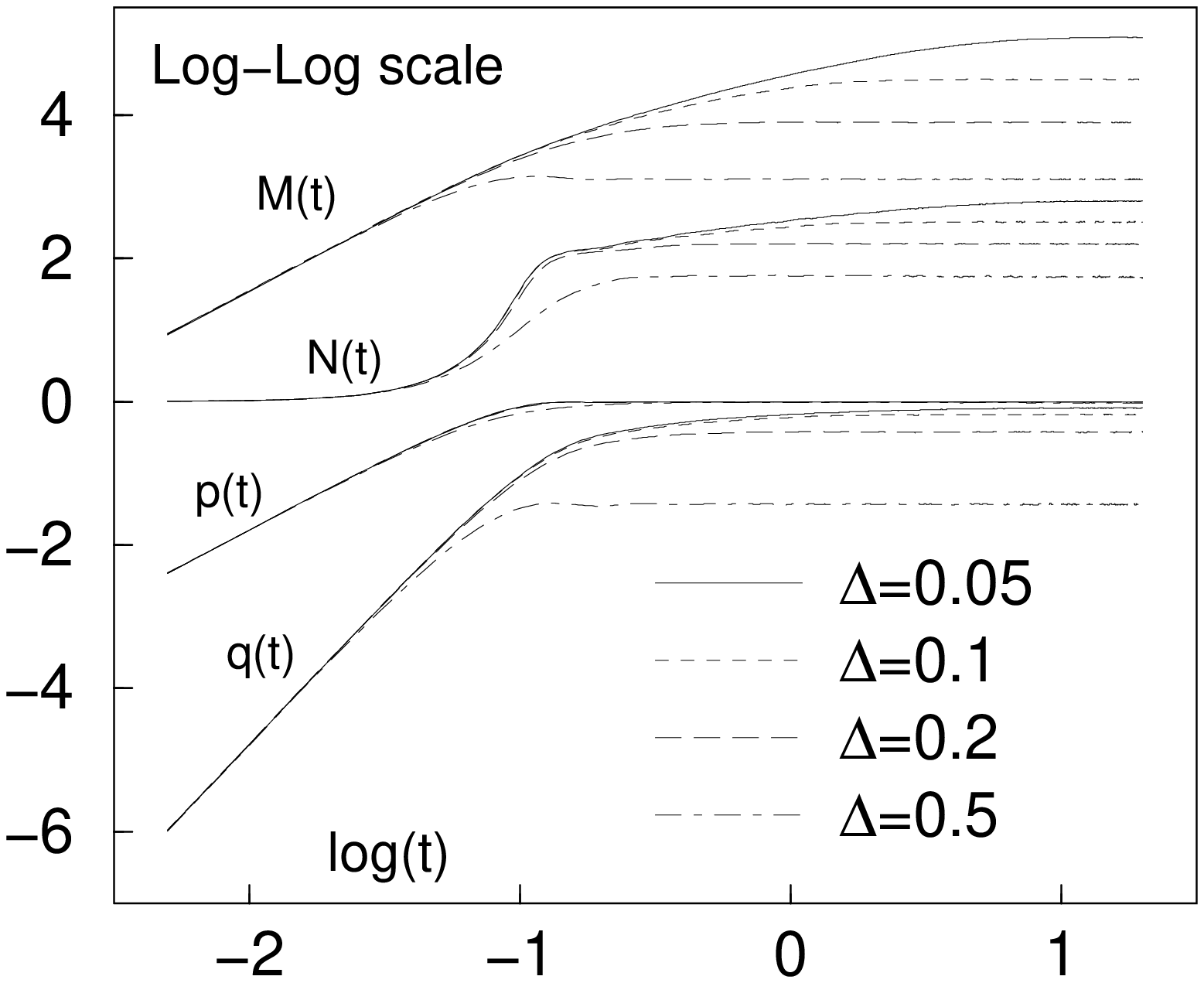}
\ \ \ \
\epsfysize=1.3in
\epsffile{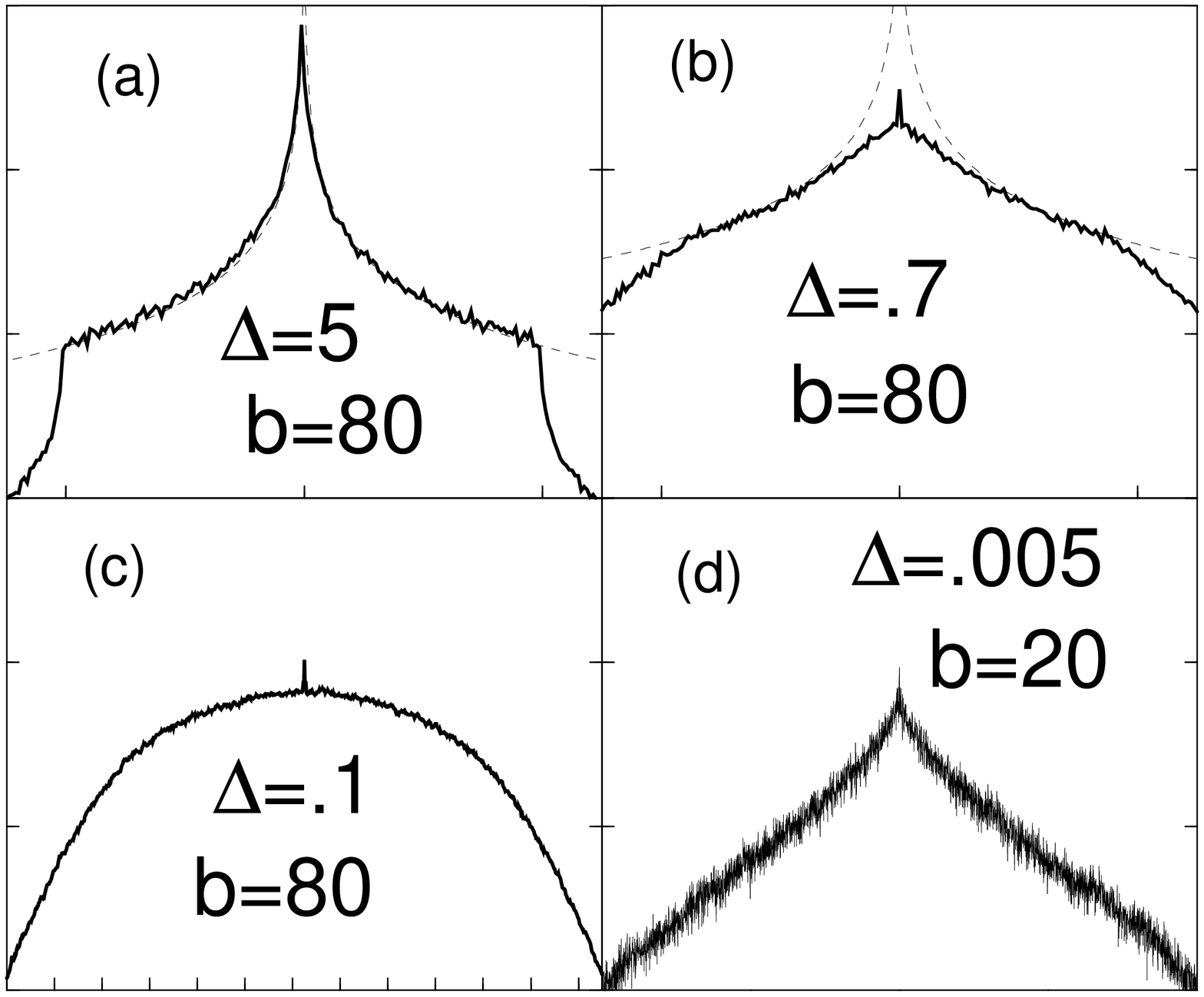}

\noindent
{\footnotesize {\bf FIG. 1.}
{\em Left figure:}
Representative examples for the time-evolution of the energy spreading profile.
The variance $M(t)$, the participation ratio $N(t)$, the total transition
probability $p(t)$, and the out-of-band transition probability $q(t)$ are
plotted for $\Delta{=}0.05,0.10,0.20,0.50$ and $b{=}80$.
The units of energy are chosen such that $\sigma{=}1$ and the units 
of time such that $\hbar{=}1$.
{\em Right figures:} Saturation profiles for various regimes: 
{\bf (a)} Standard perturbative regime; 
{\bf (b)} Wigner regime; 
{\bf (c)} Ergodic regime; 
{\bf (d)} Localization regime. 
The distance between the tick-marks on the horizontal
axis of (a)-(c) is $b$. In (d) the full scale is $|n{-}m|{<}2000$.
The full scale of the vertical log-axis is ${-}15{<}\ln(P){<}0$.
Note that the $n{=}m$ term is factor 3 larger compared with
its immediate vicinity \cite{kottos}. In (a) and (b) the
$1/(n{-}m)^2$ behavior of the (in-band) tail is fitted by dashed
lines. Notice the appearance of a core region in (b), indicated by
the `flattening' of the profile for $|n{-}m|{<}20$. 
The in-band profile in (b) corresponds to a Lorentzian 
with a very high accuracy.}
\end{figure}

In the {\em standard perturbative regime} each
eigenstate of ${\cal H}$ is localized `perturbatively' in one energy level.
Thus, for arbitrarily long times the probability is concentrated mainly
in the initial level. We can write schematically \cite{crs}
\begin{eqnarray} \label{eq4}
P_t(n|m) \approx \delta_{nm}+\mbox{Tail}(n{-}m; t)
\end{eqnarray}
where 
$\mbox{Tail}(n{-}m; t)  = ({\sigma}/{\hbar})^2
t\tilde{F}_t({(E_n{-}E_m)}/{\hbar})$ 
within the range of first order transitions 
($0<|n{-}m|<b$), and zero otherwise. 
Here  
$\tilde{F}_t(\omega)=t{\cdot}(\mbox{sinc}(\omega t/2))^2$ is the
spectral-content of a constant perturbation of duration $t$, and 
$\mbox{sinc}(x)=\sin(x)/x$. We have trivial recurrences from $n$ to $m$ 
once $t$ becomes larger than $2\pi\hbar/(E_n{-}E_m)$. 
The global crossover to quasi-periodic behavior
is marked by the Heisenberg time $t_{\tbox{H}}= 2\pi\hbar/\Delta$.
The total normalization of the tail is much less than unity at any time.

In the {\em Wigner regime}, one observes that the perturbative expression
(\ref{eq4}) is still valid for sufficiently short times $t \ll t_{\tbox{prt}}$.
Let us estimate the perturbative breaktime $t_{\tbox{prt}}$. For short times
($t<\tau_{\tbox{cl}}$) the spectral function $\tilde{F}_t(\omega)$ is very
wide compared with the bandwidth $\Delta_b$ of first-order transitions.
Consequently we can use the replacement $\tilde{F}_t(\omega)\mapsto t$ and
we get that the total transition probability is $p(t)\approx b\times
(\sigma t/\hbar)^2$. On the other hand, for $t>\tau_{\tbox{cl}}$, the spectral
function $\tilde{F}_t(\omega)$ is narrow compared with the bandwidth, and
it can be approximated by a delta-function. As a result we get $p(t)\approx
\sigma^2/(\hbar\Delta)\times t$. The condition $p(t)\sim 1$ determines 
$t_{\tbox{prt}}$ leading to:
\begin{eqnarray} \label{eq6}
t_{\tbox{prt}} =
\left\{
\matrix{
\hbar\Delta/\sigma^2 &
\mbox{for} & 1< \sigma/\Delta <\sqrt{b} \cr
\hbar/(\sqrt{b}\sigma) &
\mbox{for} & \sqrt{b} < \sigma/\Delta }
\right.
\end{eqnarray}
It should be noticed that for $\sigma\sim\Delta$ we get $t_{\tbox{prt}} \sim
t_{\tbox{H}}$.  Thus, taking recurrences into account, we come again
to the conclusion, that for $\sigma\ll\Delta$ there is no perturbative breaktime.
The {\em variance} $(\delta E(t))^2=\Delta^2\times M(t)$
of the energy distribution (\ref{eq4}) is easily calculated.
We get a ballistic-like behavior, followed by saturation,
\begin{eqnarray} \label{eq7}
\delta E(t) \approx
\left\{
\matrix{
(\delta E_{\tbox{cl}}/\tau_{\tbox{cl}}) \ t  &
\mbox{for}  & t<\tau_{\tbox{cl}} \cr
\delta E_{\tbox{cl}} &
\mbox{for}  & t>\tau_{\tbox{cl}}
}\right.
\end{eqnarray}
For $t\sim t_{\tbox{prt}}$ the tail (\ref{eq4}) becomes Lorentzian-like,
and it is characterized by a width $\hbar/t=\Gamma$. For $t>t_{\tbox{prt}}$
expression (\ref{eq4}) losses its validity, but it is obvious that the energy
cannot spread any more, since it had already acquired the saturation profile.

It should be realized that neither (\ref{eq4}), nor
the Lorentzian-like saturation profile of the Wigner regime, 
could correspond to the classical spreading profile. 
In the latter case the saturation profile is
characterized by two genuine quantum mechanical
scales ($\Gamma$, $\Delta_b$), whereas the classical
ergodic distribution is characterized by the single
energy scale $\delta E_{\tbox{cl}}$. See Fig.2. 
However, in spite of this lack of correspondence, the
variance (\ref{eq7}) behaves in a classical-like
fashion. Using the terminology of \cite{crs} 
we have here {\em restricted} rather than {\em detailed}
quantal-classical correspondence (QCC):  The quantal
$P_t(n|m)$ is definitely different from its classical
analog, but the variance $\delta E(t)$, unlike the 
higher moments of the distribution, turns out to be the same.

In the {\it ergodic regime} the time scale $\tau_{\tbox{cl}}$ becomes
larger than $t_{\tbox{prt}}$, and therefore $\tau_{\tbox{cl}}$ loses its
significance. At $t\sim t_{\tbox{prt}}$ the quantal energy-spreading just
"fills" the energy range $\Delta_b$, and we get $\delta E(t)\approx\Delta_b$.
The perturbative result (\ref{eq4}) is no longer applicable for
$t>t_{\tbox{prt}}$. However, the simplest heuristic picture turns out to
be correct. Namely, once the mechanism for ballistic-like spreading disappears
a stochastic-like behavior takes its place. The stochastic energy spreading
is similar to a random-walk process where the step size is of the order
$\Delta_b$, with transient time equals $t_{\tbox{prt}}$. Therefore we have a
diffusive behavior $\delta E(t)^2=D_{\tbox{E}}t$ where
\begin{eqnarray} \label{eq8}
D_ {\tbox{E}} \ = \ C \cdot \Delta_b^2 / t_{\tbox{prt}} \ = \
C \cdot \Delta^2 b^{\tbox{5/2}} \sigma/\hbar \ \ \propto \ \ \hbar 
\end{eqnarray}
and the numerical prefactor \cite{kottos} is $C \approx 0.85$. 
This diffusion is not of classical nature.  
The diffusion can go on as long as
$(D_{\tbox{E}} t)^{\tbox{1/2}} < \delta E_{\tbox{cl}}$, 
hence the ergodic time is 
\begin{eqnarray} \label{eq9}
t_{\tbox{erg}} \ = \ b^{\tbox{-3/2}} \ \hbar \sigma/\Delta^2
\ \propto \ 1/\hbar
\end{eqnarray}
After this ergodic time the energy spreading profile saturates to a
classical-like steady state distribution \cite{felix1}.

\begin{figure}
\epsfysize=2.5in
\epsffile{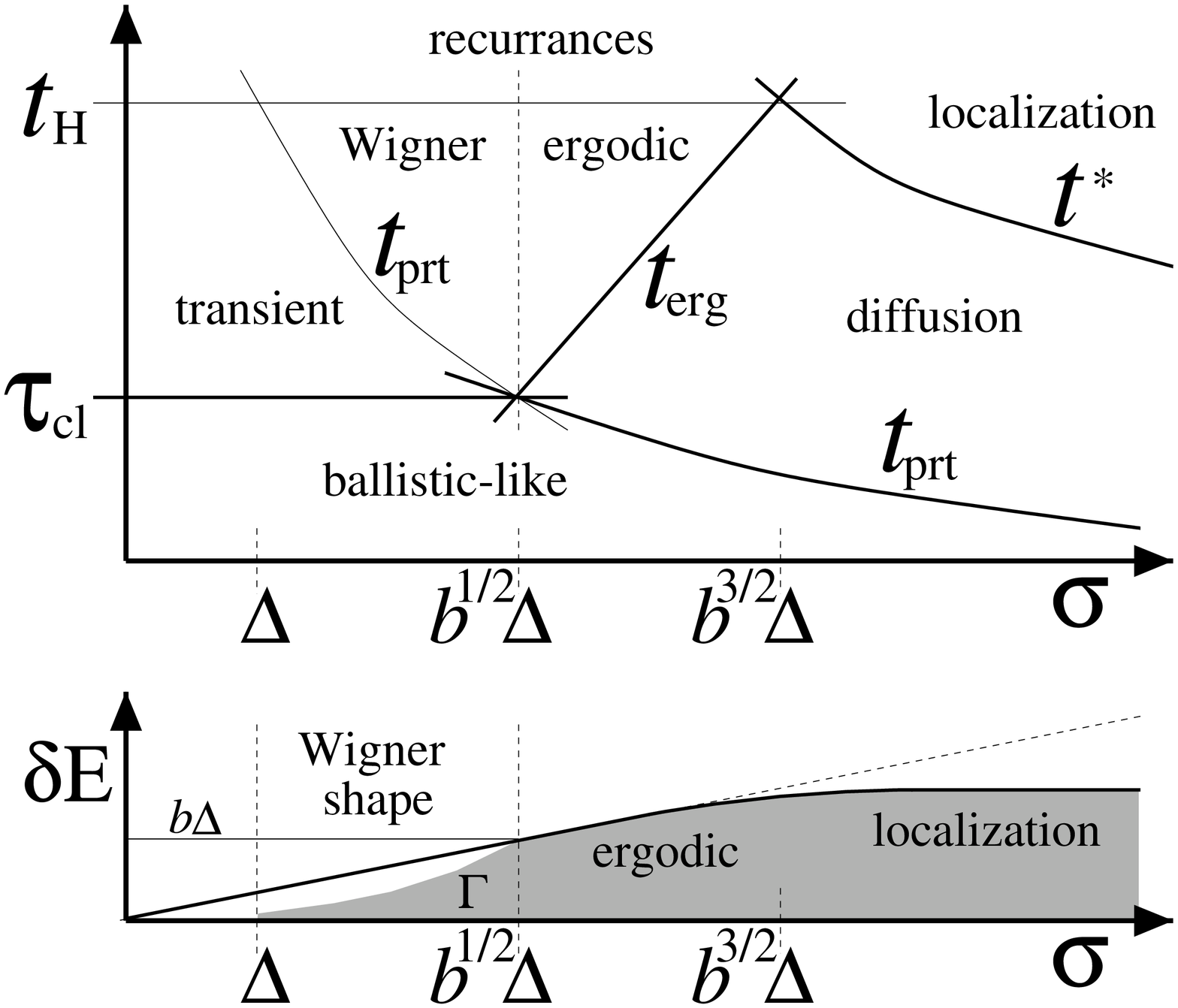}

\noindent
{\footnotesize {\bf FIG. 2.}
The {\em upper diagram} illustrates the various dynamical scenarios
which are described in the text. The flow of time is in the vertical 
direction. See \cite{brk} for closely related diagrams.
The {\em lower plot} illustrates the various energy-scales that
characterize the associated stationary distributions:
The bandwidth $\Delta_b$ is indicated by a horizontal solid line;
The width of the non-perturbative component is indicated by the
grey filling; The width of the energy shell is indicated by the
dashed line; The variance $\delta E(\infty)$ is indicated
by the bold solid line. In the localization regime
we have $\delta E(\infty)\approx\delta E_{\xi}\ll\delta E_{\tbox{cl}}$.}

\end{figure}

In the {\em localization regime} the quasi periodic nature of the dynamics
is important. The `operative' eigenstates are defined as those having a
non-negligible overlap with the initial state $m$. These eigenstates are
located within the energy shell whose width is $\delta E_{\tbox{cl}}$.
If the eigenstates are ergodic, then all of them
are 'operative', and therefore  the effective level spacing between them is
simply $\Delta_{\tbox{eff}}\approx \Delta$. However, if the eigenstates
are localized, then only $\xi$ out of them 
have a significant overlap with the initial state $m$, and therefore
the effective level spacing is 
$\Delta_{\tbox{eff}}\approx \delta E_{\tbox{cl}}/\xi$. 
The effective energy spacing $\Delta_{\tbox{eff}}$ is the relevant
energy scale for determination of the crossover to quasiperiodic behavior.
The associated time scale is $t^*=2\pi\hbar/\Delta_{\tbox{eff}}$, and
it may be either less than or equal to the Heisenberg time $t_{\tbox{H}}=
2\pi\hbar/\Delta$.
The localization regime is defined by the condition $t^*<t_{\tbox{erg}}$.
In this regime the diffusion stops before an ergodic distribution arises,
and we should get $D_{\tbox{E}} t^* \approx \delta E_{\xi}^2$.  
Inserting the definition of $t^*$
and solving for $\xi$ we obtain the well known \cite{mario,felix1} estimate
$\xi\approx b^2$. For the breaktime we obtain
\begin{eqnarray} \label{eq10}
t^* \ = \ b^{\tbox{3/2}} \hbar/\sigma
\ \propto \ (1/\hbar)^{2d{-}3}
\end{eqnarray}
Note that the localization range is $\delta E_{\xi} = \xi\Delta \propto
(1/\hbar)^{d{-}2}$. If the diffusion were of classical nature, we would get
$\delta E_{\xi} \propto (1/\hbar)^{d{-}1}$ as in the semiclassical analysis
of \cite{brk}.

\begin{figure}
\epsfysize=2.5in
\epsffile{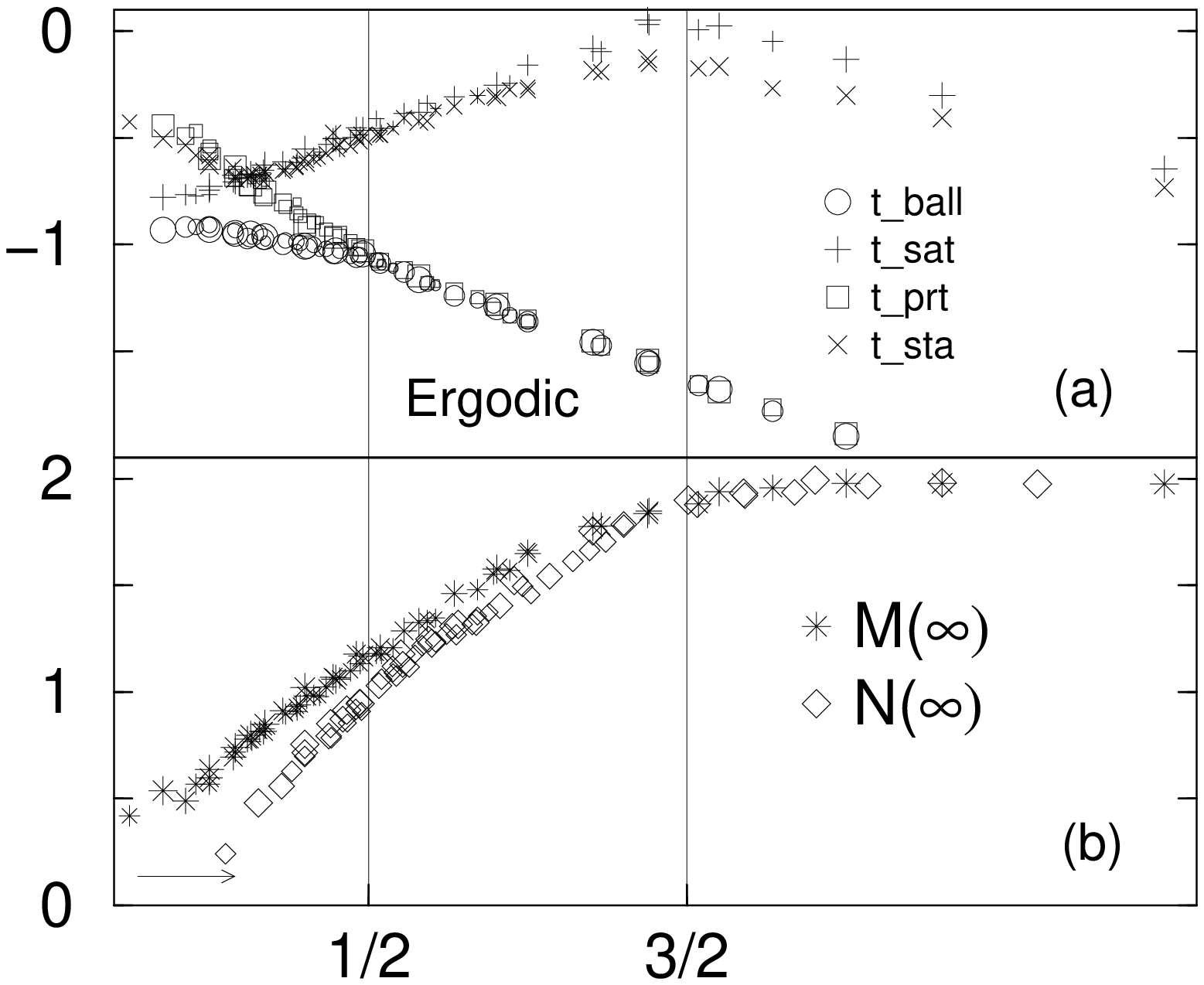}

\noindent
{\footnotesize {\bf FIG. 3.}
{\bf (a)} The times $t_{\tbox{ball}}$, $t_{\tbox{sat}}$,
$t_{\tbox{sta}}$ and $t_{\tbox{prt}}$ are numerically determined. Different
values of $b$ are distinguished by the relative size of the symbols.
The axes are  \mbox{$X=\log(\sigma/\Delta)/\log(b)$}, 
and \mbox{$Y=\log(t/t_{\tbox{H}})/\log(2\pi b)$}. Note that $Y{=}{-}1$
implies $t{=}\tau_{\tbox{cl}}$, and $Y{=}0$ implies $t{=}t_{\tbox{H}}$.
In the ergodic regime $t_{\tbox{sat}}$ departs from $t_{\tbox{bal}}$ 
and an intermediate diffusive stage appears. The saturation time 
approaches $t_{\tbox{H}}$ but eventually drops down once we enter 
into the localization regime. 
{\bf (b)} Both $\log(M(\mbox{\small $\infty$}))/\log(b)$ 
and $\log(N(\infty))/\log(b)$ are plotted versus $\log(\sigma/\Delta)/\log(b)$.
The arrow indicates a global horizontal shift of the $N(\infty)$ plot for
presentation purpose.}
\end{figure}

The various dynamical scenarios discussed above are summarized by the 
diagram of Fig.2, and can be compared with the data presented in Fig.3. 
As expected from the theoretical considerations we have in the 
Wigner regime 
$t_{\tbox{ball}} \approx t_{\tbox{sat}} \approx \tau_{\tbox{cl}}$ 
and $t_{\tbox{sta}} \approx t_{\tbox{prt}}$. In the ergodic 
regime we have as expected  
$t_{\tbox{ball}} \approx t_{\tbox{prt}} \ll \tau_{\tbox{cl}}$, 
while $t_{\tbox{sat}} \approx t_{\tbox{sta}} \equiv t_{\tbox{erg}}$.  
Thus, in the ergodic regime we have a premature 
departure of the ballistic behavior, and the appearance of an 
intermediate diffusive stage.

Our major motivation for studying WBRM model 
comes from `quantum chaos' (see introduction). 
Namely, WBRM model can be regarded as an effective model 
for the analysis of the dynamics of a `quantized' 
classically chaotic system.  
The condition to be in the regime $(\sigma/\Delta)\ll b^{\tbox{1/2}}$ 
can be cast into the form $\hbar\gg C_{\tbox{prt}}$, where 
$C_{\tbox{prt}}=\delta E_{\tbox{cl}} \tau_{\tbox{cl}}$ 
is a classical scale. In this regime the {\em perturbative} 
result Eq.(\ref{eq7}) is valid. 
The derivation of (\ref{eq7}) is not sensitive to 
the presence or the absence of subtle correlations 
between matrix elements. Therefore (\ref{eq7})
is valid in case of the `quantized' Hamiltonian, 
as well as in case of the effective WBRM model.  
Hence we may say that the applicability of an 
effective RMT approach is {\em trivial} in the 
regime $\hbar\gg C_{\tbox{prt}}$.   
In contrast to that, in the
{\em non-perturbative} regime ($\hbar\ll C_{\tbox{prt}}$),
correlations between matrix elements become important, 
and it may have implications on the dynamical behavior. 
Whether an effective RMT approach is valid 
becomes a {\em non-trivial} question in the 
non-perturbative regime.

In the regime $\hbar\gg C_{\tbox{prt}}$ we have 
restricted QCC \cite{crs}. It means that QCC holds 
{\em only} for the variance $\delta E(t)$.
Fixing all the classical parameters, including the time $t$ which is assumed
to be of the order of $\tau_{\tbox{cl}}$, we can always define a sufficient
condition $\hbar\ll C_{\tbox{SC}}$ for having detailed QCC \cite{crs}. 
Detailed QCC means that the quantal energy spreading 
profile $P_t(n|m)$ can be approximated by a classical calculation. 
The considerations that lead to the determination  of
the classical scale $C_{\tbox{SC}}$ are discussed in \cite{crs}.
We cannot give an explicit expression for $C_{\tbox{SC}}$ 
because it is a non-universal (system-specific) parameter. 
Detailed QCC implies that Eq.(\ref{eq7}) should hold again 
once the condition $\hbar\ll C_{\tbox{SC}}$ is satisfied.

For the WBRM model we have found that for $\hbar\ll C_{\tbox{prt}}$ 
there is a pre-mature departure from ballistic behavior,
followed by an intermediate diffusive behavior.
So we have a contradiction here between RMT considerations on one hand,
and QCC considerations on the other.
Thus, if the RMT approach is non-trivially valid, then it is only in a
restricted range $C_{\tbox{SC}}\ll\hbar\ll C_{\tbox{prt}}$. Outside this
regime it is either trivially valid ($\hbar\gg C_{\tbox{prt}}$) and we have
restricted QCC, or else it is not valid at all ($\hbar\ll C_{\tbox{SC}}$)
and instead we have detailed QCC. 
It may be true that in many cases the RMT considerations
are not valid for the purpose of analyzing time-dependent dynamical scenarios.
A similar situation may arise in the theory of quantum dissipation: There is
one-to-one correspondence between the regimes in Fig.5 of \cite{crs} and the
regimes that have been discussed in this Letter.


We thank the MPI f\"ur Komplexer Systeme in Dresden
for the kind hospitality during the Conference
{\it Dynamics of Complex Systems} where this work was initiated.
F.M.I acknowledges support by CONACyT (Mexico) Grants No.
26163-E and 28626-E.


\vspace*{-0.4cm}


\end{multicols}

\begin{thebibliography}{99}
\bibitem[\dag]{note} Physically, going from ${\cal H}_0$ to 
${\cal H}$ may signify a change of an external field, or 
switching on a perturbation, or sudden-change of  
effective-interaction (as in molecular dynamics).
\bibitem[\ddag]{rmrk} The numerical prefactors are chosen such that 
whenever (\ref{eq7}) applies, we have $t_{\tbox{bal}}=\tau_{\tbox{cl}}$ 
and $\delta E(\infty) = \delta E_{\tbox{cl}}$.
\end{thebibliography}
\end{document}